# Effective reduction of biofilm through photothermal therapy by gold core@shell based mesoporous silica nanoparticles

Ana García[a,b], Blanca González[a,b], Catherine Harvey[a], Isabel Izquierdo-Barba[a,b],*, and María Vallet-Regí[a,b],*

[a] *Departamento de Química en Ciencias Farmacéuticas, Unidad de Química Inorgánica y Bioinorgánica, Universidad Complutense de Madrid. Instituto de Investigación Sanitaria Hospital 12 de Octubre i+12. Plaza Ramón y Cajal s/n, 28040 Madrid, Spain.*
[b] *CIBER de Bioingeniería, Biomateriales y Nanomedicina, CIBER-BBN, Madrid, Spain.*

* Corresponding authors: Tel.: +34 91 394 1843; +34 394 1781; E-mail address: ibarba@ucm.es (Isabel Izquierdo-Barba) and vallet@ucm.es (María Vallet-Regí)

**Graphical abstract**

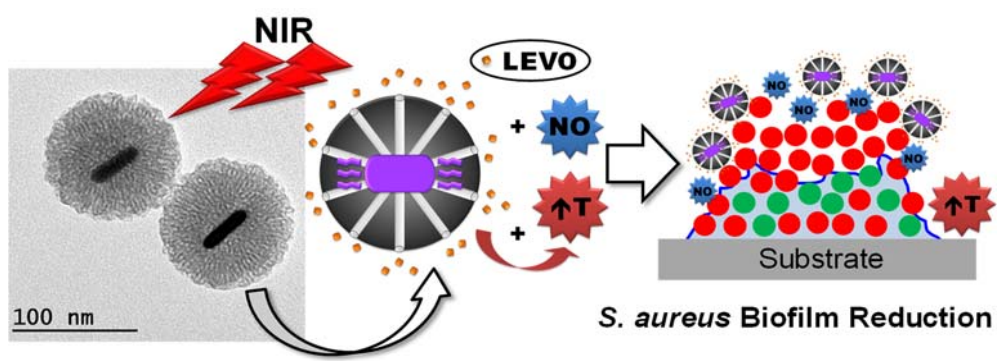

**Abstract**

Bacterial biofilms can initiate chronic infections that become difficult to eradicate. There is an unmet need for effective therapeutic strategies that control and inhibit the growth of these biofilms. Herein, light sensitive mesoporous silica nanoparticles (MSNs) with photothermal (PTT) and antimicrobial combined capabilities have been developed. These nanosystems have high therapeutic potential to affect the bacterial biofilm architecture and subsequently inhibit its growth. Nucleation of gold nanorods followed by the growth of a silica shell leads to a core@shell design (AuNR@MSN) with PTT properties. Incorporation of nitrosothiol groups (-SNO) with a heat liable linker, enables an enhanced nitric oxide release upon photothermal stimulation with near infrared radiation. Further loading of an antimicrobial molecule such as the levofloxacin (LEVO) antibiotic creates a unique nanoassembly with potential therapeutic efficacy against *Staphylococcus aureus* bacterial biofilms. A dispersion rate of the bacterial biofilm was evident when light stimuli is applied because impregnation of the nitrosothiol functionalized nanosystem with the antibiotic LEVO led to *ca.* 30% reduction but its illumination with near infrared (NIR) irradiation showed a biofilm reduction of *ca.* 90%, indicating that localized antimicrobial exposure and PTT improves the therapeutic efficacy. These findings envision the conception of near-infrared-activated nanoparticle carriers capable of combined therapy upon NIR irradiation, which enables photothermal therapy, together with the release of levofloxacin and nitric oxide to disrupt the integrity of bacterial biofilms and achieve a potent antimicrobial therapy.

# 1. Introduction

Bacterial infections are grave threat to public health and constitute the second leading cause of death worldwide [1,2]. In this sense, the infection associated with bacterial biofilm formation is hardest to deal with [3,4]. Mature biofilms are highly dynamic communities of bacteria, which are surrounded by a self-produced mucopolysaccharides layer. This layer serves as a protective barrier against the attack of antibacterial agents and the immune system, mainly because it reduces their penetrability [5,6]. As a consequence, the bacteria within the biofilm become less susceptible to antibiotics compared to individual planktonic relatives, which tends to develop the feared bacterial resistances [7]. Currently, the most widely used methods for the treatment of biofilm-related infections are conventional oral or intravenous antibiotics treatments, requiring high doses of antibiotics during long time periods [8,9]. In most cases, these treatments are ineffective in reversing new bacterial resistance and the death of the patient. In view of the above, it would be desirable to improve the penetrability of antibiotics within the biofilm in order to obtain greater therapeutic effectiveness.

In response to this challenge, several strategies for biofilm disruption and detachment were developed to degrade the protective layer and ultimately eliminate biofilm bacteria, such as amphiphilic cationic molecules [10], enzymes as proteases [11], DNase [12], and light-activated antimicrobial agents [13]. Among these last therapies, photothermal therapy (PTT) have shown a bactericidal mechanism, which differs from those of conventional antibiotic therapy. This therapy is based on the conversion of light into localized heating, mediating the strong absorption of certain metallic nanoparticles [14,15] and nanomaterials [16]. This is particularly effective in the near infrared (NIR) spectral range between 650 and 900 nm, known as the first biological window, which has been able to destroy the bacterial biofilm [17,18]. However, these therapies still represent an underestimated problem since the nonlocalized heat generally leads to serious injury of healthy tissues when eliminating biofilms. To avoid heat damage to the healthy tissue and increase local drug concentration in the biofilm, nanoscale carriers with exciting properties would be useful.

In this sense, the improvement in nanotechnology offers a new strategy to address the challenge through the development of antibiotic nanocarriers with higher efficiencies and lower side effects [19-23]. Among them, mesoporous silica nanoparticles (MSNs) constitute one of the most promising inorganic nanomaterials. MSNs exhibit outstanding features for being successfully used as smart drug delivery systems [24-30]. The main strengths of MSNs are high loading capacity, biocompatibility, easy synthesis and functionalization, robustness, and high degree of tunability regarding size, morphology and pore diameter [31]. Furthermore, due to the great versatility of MSNs design possibilities, unique stimulus-responsive nanodevices can be prepared, taking advantage of other therapeutic mechanisms such as PTT [32-34].

Nitric oxide (NO) is a free radical gaseous molecule which is endogenously produced and has an important role in intra- and extra-cellular signalling to regulate multiple functions in physiological processes [35]. In addition, delivery of exogenous NO has demonstrated potential therapeutic applications regarding cardiovascular diseases, cancer and bacterial infections among others [36]. Due to the highly reactive nature of the NO molecule, designing storage materials as vehicles with NO-releasing properties is a useful strategy for NO-based therapies, and both molecular and nanoparticle systems have been reported for fighting diseases [37-41]. NO releasing materials are usually developed by incorporating NO donor moieties in the molecules or nanoparticles. N-diazeniumdiolates and S-nitrosothiols are the most commonly used functional groups for NO delivery. S-nitrosothiols are endogenous NO donors that spontaneously release NO. S-nitrosothiols decompose via several pathways, and conditions such as light, temperature and metal ions presence can facilitate the NO release [42,43]. Regarding infection, NO has broad spectrum antibacterial activity mainly due to the generation of reactive byproducts (*e.g.*, peroxynitrite and dinitrogen trioxide) which produce an oxidative and nitrosative stress, and kill the bacteria by multiple pathways. In recent years, NO has also been identified as a key regulator of biofilm dispersal [44]. Moreover, the combination of conventional antibiotics with NO is a potential anti-biofilm strategy because bacteria becomes more susceptible to the action of the antibiotic when the NO helps with the transition from biofilm to the planktonic state [45].

This manuscript describes the design of a multifunctional hybrid organic-inorganic nanosystem with application in infection treatment. Herein, NIR-activated nanoparticles able to reduce *Staphylococcus aureus* biofilm have been developed. These nanosystems combines PTT due to the incorporation of gold nanorods into MSNs forming core@shell type nanoparticles (AuNR@MSN). Furthermore, release of antimicrobial agents (levofloxacin and nitric oxide) from a unique platform shows a combined effect against bacterial biofilm. The antibiotic levofloxacin (LEVO) is incorporated into the mesoporous structure of the nanosystems, while the nitric oxide is integrated through nitrosothiol groups (-SNO), which possess a heat liable linker, therefore enabling the stimulated NO release through NIR activation of the nanosystems. The synthesis and physico-chemical characterization of the nanosystems, the *in vitro* LEVO release study in presence and absence of NIR activation and the parameters optimization for biofilm treatment are first described. Finally, the action of these core@shell nanosystems onto preformed *S. aureus* biofilm has been tested, showing a combined effect of all their components that leads to the effective reduction of the biofilm. Figure 1 shows a schematic representation of the nanosystem and the experimental design for the antimicrobial biofilm treatment.

## 2. Experimental

### 2.1. Reagents and equipment

All reactions for the functionalization of silica surface were performed under an inert atmosphere by using standard Schlenk techniques. Solvents were dried by standard procedures and distilled immediately prior to use or bought as anhydrous solvents and kept under nitrogen. Manipulations of materials involving mesoporous silica functionalized with nitrosothiol groups were carried out in the absence of light. All glassware for the synthesis of gold nanoparticles and core@shell nanoparticles was washed with *aqua regia*, rinsed with water, washed 3-fold with Milli-Q water and dried before use.

Tetraethylorthosilicate (TEOS), cetyltrimethylammonium bromide (CTAB), 3-mercaptopropyltrimethoxysilane (MeO)$_3$Si(CH$_2$)$_3$SH 95% (MPTS), *tert*-butylnitrite (TBN), tetrachloroauric(III) acid trihydrate 99.9%, levofloxacin and Griess reactive kit were purchased from Sigma-Aldrich. 3[-Methoxy(polyetylenoxy)propyl]trimethoxysilane (PEG$_{6-9}$(CH$_2$)$_3$Si(OMe)$_3$) 90% was purchased from ABCR GmbH & Co.KG. All other chemicals (*L*-ascorbic acid, sodium borohydride, ammonium nitrate, absolute EtOH, anhydrous toluene, diethylether, NaOH, etc.) were of the highest quality commercially available and used as received. Milli-Q water (resistivity 18.2 MΩ·cm at 25 °C) was used in all experiments.

The analytical methods used to characterize the synthesized compounds were as follows: Fourier transform infrared (FTIR) spectroscopy, UV-Vis and fluorescence spectroscopy, chemical microanalyses, electrophoretic mobility measurements to calculate the values of zeta-potential (ζ), dynamic light scattering (DLS), transmission electron microscopy (TEM) and energy dispersive X-ray spectroscopy (EDS). NIR radiation was provided by a diode laser emitting at 808 nm. The equipment and conditions used are described in the Supplementary Material.

### 2.2. Materials synthesis

**AuNRs.** Gold nanorods were synthetized following a seed-mediated method and gold concentration was determined using the absorbance at 400 nm in the UV-vis spectra [33,46,47]. Briefly, 400 μL of NaBH$_4$ 1 mg/mL were added under vigorous stirring to a solution of 4.6 mL of 0.1 M CTAB and 25 μL of 0.05 M HAuCl$_4$·3H$_2$O at 30 ºC and the reaction was maintained for 30 min to get rid of the excess of NaBH$_4$. Afterwards, another solution was prepared by adding to 100 mL of 0.1 M CTAB the following in order at 30 ºC under mild stirring: 1 mL of 0.05 M HAuCl$_4$·3H$_2$O, 1.90 mL of 1 M HCl, 0.75 mL of 0.1 M ascorbic acid, 0.80 mL of 0.01 M AgNO$_3$ and, finally, 0.50 mL of the former Au seeds solution. The reaction mixture was kept undisturbed at 30 ºC for 12-16 h. Excess of reactants were then removed from the freshly prepared colloidal solution via two cycles of centrifugation, after which the AuNRs were resuspended in 0.1 M CTAB at a final gold concentration of $5 \times 10^{-3}$ M.

**AuNR@MSN.** Synthesis of core@shell nanoparticles was performed by coating the obtained AuNRs with mesoporous silica following a previously described protocol with slight modifications [32,48]. A solution of 170 mL of $6 \times 10^{-3}$ M CTAB and 75 mL of absolute EtOH was first prepared at 30 ºC and the pH value was adjusted to *ca.* 9 by adding 100 µL of NH₄OH (25%). Then, 5 mL of the AuNRs solution were poured into the solution under stirring to homogeneity. Finally, 200 µL of TEOS were added dropwise under vigorous stirring and the reaction mixture was allowed to proceed at 60 ºC overnight. The particles were collected by centrifugation, washed with EtOH and left to dry. Pore surfactant containing material (AuNR@MSN$_{surf}$) was considered to have 40% weight from the CTAB, and weight percentages of gold and silica were calculated from the EDS analysis resulting in 8.93% wt. for Au and 91.07% wt. for $SiO_2$. When necessary, the surfactant was removed from the functionalized material by heating at 80 ºC overnight a well dispersed suspension of 30 mg of the obtained solid in 30 mL of an extracting solution (NH₄NO₃ 10 g/L in EtOH/H₂O 95:5 vol). The solid was recovered by centrifuging then washed twice with H₂O and three times with EtOH and dried.

**AuNR@MSN-PEG$_{ext}$.** For the external surface functionalization of AuNR@MSNs with PEG units, pore surfactant containing material (AuNR@MSN$_{surf}$) was employed and, therefore, approximately a quarter of the specific surface area of the free-surfactant material (*ca.* 320 m$^2$/g) [33] was considered to be functionalized. Prior to surface functionalization 0.0380 g of CTAB-containing AuNR@MSN$_{surf}$ (*i.e.*, 0.0228 g AuNR@MSN and 0.0208 g MSN) were dehydrated at 80 ºC, under vacuum for 3 h, and subsequently redispersed under an inert atmosphere in dry toluene (10 mL). A solution of PEG$_{6-9}$(CH$_2$)$_3$Si(OMe)$_3$ (2.6 mg, 10% exc.) in 1 mL of dry toluene was added to the vigorously stirred suspension of the AuNR@MSN$_{surf}$ and the mixture was heated to 100 ºC overnight in the dark. The reaction mixture was centrifuged at 11000 rpm for 15 min and the obtained solid was washed twice with toluene and three times with EtOH and finally dried. The surfactant was removed from the functionalized material by heating at 80 ºC overnight a well dispersed suspension of 30 mg of the obtained solid in 30 mL of an extracting solution (NH₄NO₃ 10 g/L in EtOH/H₂O 95:5 vol). The solid was recovered by centrifuging then washed twice with H₂O and three times with EtOH and dried.

**AuNR@MSN-SH.** For the functionalization of the 100% of the specific surface area of the core@shell nanoparticles, 0.020 g of the surfactant extracted AuNR@MSN were dehydrated at 80 ºC, under vacuum for 3 h, and subsequently redispersed under an inert atmosphere in dry toluene (5 mL). A solution of (MeO)$_3$Si(CH$_2$)$_3$SH (3.7 mg, 10% exc.) in 1 mL of dry toluene was added to the vigorously stirred suspension of the AuNR@MSN and the mixture was heated to 100 ºC overnight in the dark. The reaction mixture was centrifuged at 11000 rpm for 15 min and the obtained solid was washed twice with toluene and three times with EtOH and finally dried.

**AuNR@MSN-SNO.** AuNR@MSN-SH material (0.0148 g) was resuspended in 5 mL of EtOH/toluene (4:1 vol) and *tert*-butylnitrite (100 μL) was then added. The reaction mixture was stirred at 20 ºC overnight in the absence of light. The solid was recovered by centrifuging at 11000 rpm and 4 ºC and washed with cold EtOH, $H_2O$, EtOH and diethyl ether. Drying was performed under a $N_2$ flow and the nanosystem was stored at –20 ºC under $N_2$ in the darkness.

**2.3. Levofloxacin loading and release assays.**

AuNR@MSN, AuNR@MSN-PEG$_{ext}$ and AuNR@MSN-SNO were loaded with the antibiotic levofloxacin following the same procedure for all of them to obtain AuNR@MSN+LEVO, AuNR@MSN-PEG$_{ext}$+LEVO and AuNR@MSN-SNO+LEVO nanosystems, respectively. Briefly, 6 mg of sample were soaked in 2 mL of 0.016 M LEVO solution in dichloromethane. The obtained suspension was magnetically stirred at 5 ºC for 6 h and dark conditions. After that, the suspension was centrifuged at 4 ºC, washed twice with cold EtOH to remove the cargo adsorbed on the external surface and dried under a nitrogen stream at low temperature.

*In vial* LEVO release assays were performed with the AuNR@MSN-PEG$_{ext}$-LEVO nanosystem by soaking the nanoparticles in phosphate buffered saline solution (PBS 1×) at physiological conditions (37 ºC, pH = 7.4). Cumulative release of LEVO in the medium was determined through fluorescence spectrometry (BiotekPowerwave XS spectrofluorimeter, version 1.00.14 of the Gen5 program, with $\lambda_{ex}$ = 292 nm and $\lambda_{em}$ = 494 nm). For this purpose, AuNR@MSN-PEG$_{ext}$-LEVO sample was suspended in PBS 1× (2.5 mg/mL) and 100 μL of nanoparticle suspension was deposited in Transwell® chambers. Transwell® chambers have a permeable support that allows the released LEVO molecules diffuse into the culture plate, while the nanoparticles are held back. The well plates were filled with 900 μL of PBS 1×, so the final concentration obtained was 0.25 mg/mL. The systems were kept at 37 ºC in an orbital shaker (100 rpm) refreshing the medium for each time of liquid withdrawing. The samples with NIR treatment were irradiated for 10 minutes (808 nm, 1 W/cm$^2$) before liquid withdrawing. Calibration line for LEVO concentration was calculated for a concentration range of 0-3 μg/mL obtaining a regression coefficient 0.997.

**2.4. Microbiological assays.** *Staphylococcus aureus* (*S. aureus*, ATCC 29213 laboratory strain) was used as Gram-positive bacteria. Bacteria culture was carried out by inoculation in Todd Hewitt Broth (THB; Sigma-Aldrich) and incubated at 37 ºC with orbital shaking at 100 rpm to obtain an adequate concentration. Bacteria concentration was determined by spectrophotometry using a visible spectrophotometer (Photoanalyzer D-105, Dinko instruments).

**Biofilm growth.** *S. aureus* biofilms were developed into 24-well plates (P-24, CULTEK) for one day at 37 ºC under orbital stirring at 100 rpm. One free row between samples was left in the P-24 to avoid radiation in adjacent samples during the NIR treatment. The concentration of *S. aureus* bacteria was $10^8$ CFUs per mL and the medium used was THB supplemented with 4% sucrose to promote a robust biofilm formation. After 24 h, each well was gently washed twice with 1 mL of PBS 1× buffer solution under aseptic conditions to eliminate medium and unbound bacteria. The generated biofilms could be visually observed on the bottom of wells.

**Antimicrobial effect of AuNR@MSN nanosystems against *S. aureus* biofilms.** Quantitative antibiofilm assays were carried out by calculating the reduction of CFU/mL. The previously developed *S. aureus* biofilms were first covered with 500 μL of THB medium and then with another 500 μL of fresh AuNR@MSN materials suspensions at 100 μg/mL. A 500 μg/mL stock solution was first prepared by suspending the materials in 1 mL of PBS 1× and then diluting with THB to 100 μg/mL. The final volume per well was 1 mL obtaining AuNR@MSN material concentrations of 50 μg/mL. Biofilm controls without nanosystems were used and the experiments were performed in triplicate. After 90 minutes of incubation at 37 ºC and 100 rpm, samples were exposed to NIR laser treatment (808 nm, 1 W/cm$^2$, 10 min). Subsequently, the plates were incubated for another 90 minutes (37 ºC, 100 rpm) and then received a second NIR irradiation in the same conditions. For each nanosystem, control samples were subjected to the same experimental conditions without the NIR laser treatment. All well plates, with or without NIR treatment, were incubated at 37 ºC during 24 hours with orbital shaking. After incubation, the plates were washed twice with sterile PBS 1× and sonication was applied for 10 minutes in a low-power bath sonicator (Selecta, Spain) to break and disperse the biofilm in a total volume of 1 mL of PBS 1×. The sonicated biofilms were diluted to 1:100 (once) and 1:10 (3 times) in a final volume of 1 mL. Quantification of bacteria was carried out using the drop plate method [49]. Seven drops of each solution were inoculated in Tryptic Soy Agar (TSA, Sigma Aldrich) plates which were incubated for 24 h at 37 ºC. The mean count of the 7 drops of each dilution was calculated, and then, the average counting for all dilutions was calculated following the procedure described in reference [50]. In addition, a preliminary assay to confirm the PTT local effect onto the biofilm has been performed by confocal microscopy. For this propose, 50 μg/mL of AuNR@MSN sample was incubated onto the mature biofilm during 90 min and after that time the PTT previously described was carried out. The resulting biofilm was washed three times with sterile PBS 1× and then 3 μL/mL of Live/Dead® Bacterial Viability Kit (Backlight$^{TM}$) was added to stain live bacteria in green and dead bacteria in red. Then, 5 μL/mL of calcofluor solution was also added to stain the mucopolysaccharides of the biofilm (extracellular matrix) in blue. Both reactants were incubated for 15 min at room temperature. Controls containing biofilm bacteria were also performed. The samples were examined in an Olympus FV1200 confocal microscope.

## 3. Results and Discussion

The synthesis of nanosystems is depicted in Figure 2. In addition to the core@shell nanosystem functionalized with the nitrosothiol groups, we have prepared a compound functionalized with a short PEG oligomer in the external surface as a model to carry out the loading and release studies of levofloxacin antibiotic. This model compound has also been used to adjust the conditions of nanoparticles concentration, laser power and radiation time to optimize the reached temperature for the PPT effect. The preparation of this compound offers us the advantage of being able to avoid the necessary precautions required to work with derivatives that contain nitrosothiol groups, which have some instability in the presence of light and temperature. In the first step, gold nanorods were prepared following a seed mediated method in aqueous solution which leads to CTAB-stabilized AuNRs [33,46,47]. UV-Vis-NIR spectrophotometry analysis of the synthetized AuNRs shows an intense absorbance around 800 nm due to the localized surface plasmon resonances (Figure 3A). This spectral range corresponds to the near infrared wavelength of 808 nm emitted by the diode laser used in our experiments aimed at working in the biological window. A representative TEM image of the AuNRs is shown in Figure 3B, which confirms the monodispersity of the sample and the rod-like morphology of the nanoparticles, having an average length and width of $52.4 \pm 4.9$ nm and $14.4 \pm 3.4$ nm.

The mesoporous silica shell is formed through hydrolysis and condensation of TEOS following a template growth using the CTAB-stabilized AuNRs as nucleation sites in the presence of cationic surfactant CTAB as structure directing agent [51]. In addition, a basic media following a modified Stöber method is used [32]. Therefore, the AuNRs were then coated with a mesoporous silica shell in one-step, obtaining core@shell nanoparticles having a spherical morphology with uniform diameter of *ca.* 90 nm as shown in the TEM micrographs (Figures 3C and 3D). This result is also verified by DLS measurements of the hydrodynamic size of the materials (Table 1), where a monomodal and narrow distribution from *ca.* 30 to 110 nm was found. In addition, TEM images revealed radially oriented mesochannels with pore diameters of approximately 2 nm. EDS analysis registered at low magnifications of the core@shell materials showed atomic percentages for gold and silicon of 2.9% Au and 97.1% Si, representing a 0.004 molar ratio of Au/Si in the nanosystems.

Functionalization of AuNR@MSNs with alkoxysilanes was performed through the post-synthesis method in the absence of water molecules [52,53]. To attach the PEG oligomer onto the external surface of the AuNR@MSNs, the post-grafting reaction of the $PEG_{6-9}(CH_2)_3Si(OMe)_3$ was conducted on the as-synthetized AuNR@MSNs containing the surfactant filling the pores to avoid diffusion of the alkoxysilane into the mesoporous structure [54,55]. However, to achieve maximum NO donors, the functionalization with mercaptopropyltrimethoxysilane was performed on the surfactant extracted AuNR@MSNs, so the incorporation of –SH groups is achieved also in the inner mesoporous silica surface. In both cases, the required amount of functionalizing agent was calculated for a 100% nominal degree of surface functionalisation plus a 10% excess. We took into account the specific surface area of this kind of materials [33] of which approximately a quarter was estimated to correspond to the external surface for the PEG functionalization [56]. The value for the average surface concentration of Si-OH groups in the amorphous silica materials used was 4.9 OH/nm$^2$, as estimated by Zhuravlev [57]. The stoichiometry considered for the condensation reaction between the free silanol groups of the silica exterior surface and the alkoxysilane-functionalised derivatives was a molar ratio of three Si-OH groups to one R-Si(OEt)$_3$ moiety [56]. The thiolated AuNR@MSN-SH were provided with the –SNO groups by reacting the –SH groups with a nitrosating agent such as *tert*-butylnitrite [58,59].

Changes in the zeta-potential ($\zeta$) values of the AuNR@MSNs were used to follow the incorporation of the different functional groups in the nanosystems, taking into account the acid-base equilibrium of the different functional groups over the nanoparticle silica surface in water. As shown in Table 1, the direct grafting of the PEG oligomer produced a slightly less negative $\zeta$-potential value compared to the bare AuNR@MSNs. This fact can be ascribed to the consumption of –SiO$^-$ groups of the silica surface (Eq. 1) in the condensation reaction. However, the introduction of the mercaptopropylsilane hardly changes the value of the $\zeta$-potential with respect to the non-functionalized silica surface, due to the presence of thiol groups (-SH) that generates thiolates (-S$^-$) in their acid-base equilibrium (Eq. 2). Once nitrosation of the thiol groups occurs, the $\zeta$-potential increases due to the conversion of thiol groups to nitrosothiol groups that does not undergo acid base equilibrium in the water media.

$$\text{R-Si-OH} + \text{H}_2\text{O} \rightleftharpoons \text{R-SiO}^- + \text{H}_3\text{O}^+ \qquad \text{p}K_a \approx 6.8 \qquad \text{Eq. 1} \quad [29,60]$$

$$\text{R-SH} + \text{H}_2\text{O} \rightleftharpoons \text{R-S}^- + \text{H}_3\text{O}^+ \qquad \text{p}K_a \approx 10.6 \qquad \text{Eq. 2} \quad [61]$$

Results of the hydrodynamic size distributions obtained by DLS in water media (Figure S1) show very similar profiles with a monomodal and narrow distribution for all the samples. Moreover, except for the bare AuNR@MSN, hydrodynamic size distributions are centered around 100 nm for all the functionalized nanosystems. Values of the hydrodynamic diameter obtained by DLS measurements in water are also in accordance with changes in the $\zeta$-potential when comparing the bare AuNR@MSNs and the functionalized nanosystems (see Table 1 and Figure S1). For the bare AuNR@MSNs the maximum of the size distribution is found around 68 nm, which is shifted to *ca.* 91 nm when the PEG oligomers are attached to the external surface. On the one hand, hydrodynamic size increases due to the presence of the PEG chains, but it also affects the increase of the ζ-potential towards the zone of colloidal instability, which causes a greater number of aggregates and the displacement of the maximum of size distribution towards higher values. When the mercaptopropyl is attached, the hydrodynamic size increases a smaller amount, to *ca.* 79 nm, because, although organic chains are also introduced on the surface of the nanoparticles, the ζ-potential of this nanosystems still has values of –20 mV, very similar to the unfunctionalized material, *i.e.*, in the same zone of colloidal stability. When the –SNO groups are introduced over the –SH groups, the maximum of the size distribution moves increasing up to *ca.* 106 nm reflecting an increase in the ζ-potential towards the zone of colloidal instability, up to –12 mV.

The incorporation of the alkoxysilane derivatives on the AuNR@MSNs was also followed by quantification of the organic content of the nanosystems by elemental chemical analyses (Table 2). The results confirm the expected increase of the C% content after functionalization of the AuNR@MSNs with the PEG-alkoxysilane or the mercaptopropylsilane, keeping the values of H% and N% practically constant. In the case of functionalization with mercaptopropylsilane, S% increases confirming the presence of –SH groups. The values recorded for the nitrosated sample are very similar to the thiolated sample. In this case, no increase in the N percentage was expected because of the intrinsic thermal instability of the –SNO group which must decompose in the prechamber oven of the microanalyzer.

Infrared spectroscopy was as well used to confirm the correct functionalization of AuNR@MSNs with thiol and nitrosothiol groups. FTIR spectra of the core@shell nanoparticles are shown in Figure S2 (Supplementary Material). The FTIR spectrum of AuNR@MSN$_{surf}$ displays two characteristic bands *ca.* 2920 and 2850 cm$^{-1}$ assigned to ν(C-H) of -CH$_2$- chains of CTAB that disappear in the extracted AuNR@MSN sample, therefore confirming the correct surfactant removal. AuNR@MSN-SH and AuNR@MSN-SNO spectra show the emergence of a new band at *ca.* 2975 cm$^{-1}$ related to aliphatic -CH$_2$- vibrations which confirms the proper grafting of propyl groups introduced during the functionalization procedure with mercaptopropylsilane functionalization agent. The presence of characteristics bands of the -S-N=O group are observed in the AuNR@MSN-SNO spectrum which displays a weak peak *ca.* 1450 cm$^{-1}$ and a shoulder *ca.* 800 cm$^{-1}$ assigned to vibrations of -S=N and -N=O, respectively, therefore confirming the successful introduction of the nitrosothiol groups [39,58]. All samples display typical bands of silica materials due to the Si-O stretching vibrations at *ca.* 1040 and 790 cm$^{-1}$, respectively, and the bending Si-O-Si strength at 430 cm$^{-1}$.

The last step in the preparation of the core@shell nanosystems was the loading into the mesoporous structure of a wide spectrum antibiotic such as levofloxacin. The LEVO loading was performed through an impregnation method by soaking the nanosystems at 5 °C in a 0.016 M LEVO solution prepared in dichloromethane. LEVO is readily soluble in dichloromethane even at low temperature which helps the preservation of the -SNO group during the loading procedure. Moreover, the use of a concentration double than the usual 0.008 M, allows reducing loading times to just 6 hours instead of 16 or 24 h.

Once the different materials have been characterized, the next step was to evaluate the effect of the light-triggered response of the nanosystems and its potential in PTT. To this end, the temperature increase of the medium was first evaluated *in vial* for the AuNR@MSN-PEG$_{ext}$ sample. To optimize the temperature reached with the photothermal effect of the nanosystems we have analysed different parameters such as nanoparticles concentration (10-100 μg/mL), laser power (0.5 and 1 W/cm$^2$) and radiation time (5 and 10 min) using glass covers as substrate (Figure 4A). The obtained results show a temperature increase as a function of concentration, finding an optimal photothermal effect for the concentration range between 50-80 μg/mL. For higher concentrations such as 100 μg/mL, a temperature of *ca.* 50 ºC is reached, which falls above the safe range of hyperthermia [62]. The *in vial* LEVO release in the presence and absence of NIR radiation has been also studied. The amount of LEVO loaded in AuNR@MSN-PEG$_{ext}$+LEVO has been determined as 25.8 μg/mg of material from the difference in %N between unloaded and drug loaded samples obtained by elemental chemical analyses (Table 2). Figure 4B shows the cumulative LEVO concentration as a function of time for the AuNR@MSN-PEG$_{ext}$+LEVO sample, where similar profiles were recorded regardless of the application of the NIR. In both cases, LEVO release profiles display typical-diffusion kinetic from mesoporous matrices. These results suggest that, under the conditions applied, LEVO release is controlled by its interaction with the silica surface and is not affected with the temperature increase reached due to NIR irradiation [23]. Therefore, the interaction of the LEVO with the surface silanol groups, mainly ascribed to hydrogen bonds at pH 7.4, has a greater contribution than the possible photothermal-promoted drug release effect in the temperature range reached after NIR irradiation [23,63,64].

The antimicrobial capability of these nanosystems has been assayed onto mature *S. aureus* biofilm. The effect of photothermal therapy (PTT) after two cycles of NIR laser irradiation was evaluated in presence or not of each of the antimicrobial agents, LEVO or NO, or a combination of both. The -

SNO groups incorporated into the silica surface of the core@shell nanosystems are able to release nitric oxide molecules. It was also evaluated the antibacterial effect in the absence of the NIR irradiation attributed to the individual release of LEVO or NO or their combined effect. With this aim, preformed *S. aureus* biofilms were incubated with suspensions of the different nanosystems in a 50 μg/mL final concentration at 37 ºC for 90 minutes before being subjected to a first treatment with the NIR laser (808 nm, 1 W/cm$^2$ during 10 minutes). The same procedure was repeated after another 90 minutes of incubation, and finally all samples (with or without PTT treatment) were incubated for 24 hours at 37 ºC with orbital shaking. The results obtained are displayed in Figure 5 that quantitatively shows the reduction of cell viability expressed as CFU/mL in the biofilm after applying the different conditions.

For AuNR@MSN nanosystem, a similar effect was observed with and without PTT treatment, taking place a reduction of CFU/mL in the biofilm of *ca.* 20%. These results are in agreement with a confocal microscopy study where hardly any differences were observed in the presence or absence of treatment, showing in both cases a blanket of green bacteria over the entire biofilm (A and B images in Figure S3). The difference for the sample subjected to PTT treatment is that several ablation nuclei in red can be observed throughout the biofilm.

However, when AuNR@MSN mesopores are loaded with LEVO the improvement in bactericidal activity is significant, increasing to 38.4% for the AuNR@MSN+LEVO nanosystem without PTT treatment and 66.4% for the system subjected to NIR laser irradiation. Undoubtedly, this increase in reduction of CFU/mL in the biofilm in the samples containing LEVO must be due to the antibiotic presence. However, the higher antimicrobial effect after NIR treatment could be attributed to the increase of local temperature by the action of NIR treatment which also provokes ablation nuclei throughout the biofilm matrix, which may help the LEVO action thus increasing the bactericidal efficiency of the nanosystem. The confocal microscopy image of this sample AuNR@MSN+LEVO after the NIR treatment shows a significant decrease in both the live bacteria and the mucopolysaccharide matrix (Figure S3C).

The presence of nitrosothiol groups anchored to the silica surface of the nanoparticles promoted somewhat higher antibacterial efficacy compared to bare AuNR@MSN nanosystem. The -SNO group decomposes and nitric oxide is released into medium. Since the -SNO group is thermal labile, we checked that the NO release from our nanosystems increases with the temperature of *ca.* 45 ºC reached in our experiment after NIR irradiation (see Figure S4) [42,43]. In this sense, when AuNR@MSN-SNO system was heated by the PTT action, the bactericidal activity of the AuNR@MSN-SNO was slightly increased from 39.4 to 45.4%. In fact, the confocal microscopy image (Figure S3D) shows a slight reduction of the mucopolysaccharide biofilm matrix with an increase in dead bacteria coexisting with live bacteria, in agreement with the bacteria quantification that shows a lower effect in CFU/mL reduction than the LEVO containing sample AuNR@MSN+LEVO. This difference can be attributed to the more potent effect of the levofloxacin antibiotic [23].

Finally, the LEVO incorporation to the system (AuNR@MSN-SNO+LEVO) does not improve the capacity of the free-LEVO AuNR@MSN nanosystem, obtaining only 31.4% reduction of CFU/mL in biofilm in the absence of NIR laser irradiation. However, when AuNR@MSN-SNO+LEVO is treated with NIR irradiation the reduction of CFU/mL in biofilm was 88%, suffering a quasi-complete bacteria eradication. The analysis by confocal microscopy displays a synergistic effect for this situation (Figure S3E), presenting that almost all of the live bacteria and mucopolysaccharide zones have disappeared, and with only a few isolated live bacteria showing up. This enhanced effect of LEVO and NO together could be attributed to the matrix destabilization provoked by the hyperthermia effect. This quasi biofilm destruction could be attributed to a three-factor combination therapy: PTT treatment, which is responsible for local temperature rise, NO enhanced release and LEVO release. Therefore, heat from PPT treatment would produce biofilm dispersal making bacteria more susceptible to the action of the released NO and LEVO.

## 4. Conclusions

Multifunctional thermosensitive mesoporous silica nanoparticles (MSNs) with photothermal and antimicrobial cooperative capabilities against mature *Staphylococcus aureus* biofilm have been designed. This nanosystem possesses photothermal therapy due to the incorporation of gold nanorods into MSNs forming core@shell-type nanoparticles (AuNR@MSN). Furthermore, the increase of temperature upon near infrared stimuli and the release of antimicrobial agents (levofloxacin and nitric oxide) from a unique platform shows combined effect against bacterial biofilm. The levofloxacin is incorporated into the mesoporous channels of the nanosystems, while the nitric oxide is integrated through a heat liable linker such as nitrosothiol groups (-SNO), therefore also enabling a higher NO release through NIR activation. This nanosystem has high therapeutic potential to affect the bacterial biofilm architecture and subsequently inhibit its growth. Its potent antimicrobial activity can be attributed to the threefold effect of photothermal therapy since effectively disrupt the integrity of *S. aureus* biofilm, together with the enhanced nitric oxide agent release and the bactericidal activity of levofloxacin, which is augmented in a dispersed biofilm, hence conceptualizing a compelling nanoplatform for antimicrobial therapy.

**Conflicts of interest**

The authors declare that they have no known competing financial interests or personal relationships that could have appeared to influence the work reported in this paper.

**Acknowledgement**

Authors acknowledge funding from the European Research Council (Advanced Grant VERDI; ERC-2015-AdG Proposal No. 694160) and the Ministerio de Ciencia e Innovación (PID2020-117091RB-I00 grant).

**References**

[1] Global Antimicrobial Resistance and Use Surveillance System (GLASS) Report: 2021. WHO, Global report. Available online: https://www.who.int/publications/i/item/9789240027336 (accessed on 24 June 2021).

[2] J. O'Neill, *Rev. Antimicrob. Resist.* 2016, **1**, e1600525. Tackling drug-resistant infections globally: final report and recommendations.

[3] D. Davies, Nat. Rev. Drug Discov. 2 (2003) 114-122. https://doi.org/10.1038/nrd1008

[4] L. Hall-Stoodley, J.W. Costerton, P. Stoodley, Nat. Rev. Microbiol. 2 (2004) 95-108. https://doi.org/10.1038/nrmicro821

[5] H.-C. Flemming, J. Wingender, Nat. Rev. Microbiol. 8 (2010) 623-633. https://doi.org/10.1038/nrmicro2415

[6] I. Sutherland, Trends Microbiol. 9 (2001) 222-227. https://doi.org/10.1016/S0966-842X(01)02012-1

[7] S.B. Levy, B. Marshall, Nat. Med. 10 (2004) S122-S129. https://doi.org/10.1038/nm1145

[8] P.S. Stewart, J.W. Costerton, Lancet 358 (2001) 135-138. https://doi.org/10.1016/S0140-6736(01)05321-1

[9] L.R. Hoffman, D.A. D'Argenio, M.J. MacCoss, Z. Zhang, R.A. Jones, S.I. Miller, Nature 436 (2005) 1171-1175. https://doi.org/10.1038/nature03912

[10] T.E. Angelini, M. Roper, R. Kolter, D.A. Weitz, M.P. Brenner, Proc. Natl. Acad. Sci. 106 (2009) 18109-18113. https://doi.org/10.1073/pnas.0905890106

[11] P. Zhang, S. Li, H. Chen, X. Wang, L. Liu, F. Lv, S. Wang, ACS Appl. Mater. Interfaces 9 (2017) 16933-16938. https://doi.org/10.1021/acsami.7b05227

[12] C.B. Whitchurch, T. Tolker-Nielsen, P.C. Ragas, J.S. Mattick, Science 295 (2002) 1487. DOI:10.1126/science.295.5559.1487

[13] X. Dai, Y. Zhao, Y. Yu, X. Chen, X. Wei, X. Zhang, C. Li, ACS Appl. Mater. Interfaces 9 (2017) 30470-30479. https://doi.org/10.1021/acsami.7b09638

[14] Y. Li, T. Wen, R. Zhao, X. Liu, T. Ji, H. Wang, X. Shi, J. Shi, J. Wei, Y. Zhao, X. Wu, G. Nie, ACS Nano 8 (2014) 11529-11542. https://doi.org/10.1021/nn5047647

[15] Y. Wang, K.C.L. Black, H. Luehmann, W. Li, Y. Zhang, X. Cai, D. Wan, S. Y. Liu, M. Li, P. Kim, Z.-Y. Li, L. V Wang, Y. Liu, Y. Xia, ACS Nano 7 (2013) 2068-2077. https://doi.org/10.1021/nn304332s

[16] J.-W. Xu, K. Yao, Z.-K. Xu, Nanoscale 11 (2019) 8680-8691. https://doi.org/10.1039/C9NR01833F

[17] R.S. Norman, J.W. Stone, A. Gole, C.J. Murphy, T.L. Sabo-Attwood, Nano Lett. 8 (2008) 302-306. https://doi.org/10.1021/nl0727056

[18] N. Levi-Polyachenko, C. Young, C. MacNeill, A. Braden, L. Argenta, S. Reid, Int. J. Hyperthermia 30 (2014) 490-501. https://doi.org/10.3109/02656736.2014.966790

[19] Y. Liu, H.J. Busscher, B. Zhao, Y. Li, Z. Zhang, H.C. van der Mei, Y. Ren, L. Shi, ACS Nano 10 (2016) 4779-4789. https://doi.org/10.1021/acsnano.6b01370

[20] M. Vallet-Regí, M. Colilla, I. Izquierdo-Barba, The Enzymes 44 (2018) 35-59. https://doi.org/10.1016/bs.enz.2018.08.001

[21] M. Vallet-Regí, B. González, I. Izquierdo-Barba, Int. J. Mol. Sci. 20 (2019) 3806. https://doi.org/10.3390/ijms20153806

[22] M. Vallet-Regí, D. Lozano, B. González, I. Izquierdo-Barba, Adv. Healthcare Mater. 9 (2020) 2000310. https://doi.org/10.1002/adhm.202000310

[23] A. Aguilar-Colomer, M. Colilla, I. Izquierdo-Barba, C. Jiménez-Jiménez, I. Mahillo, J. Esteban, M. Vallet-Regí, Microporous Mesoporous Mater. 311 (2021) 110681. https://doi.org/10.1016/j.micromeso.2020.110681

[24] M. Vallet-Regí, A. Rámila, R. P. del Real, Chem. Mater. 13 (2001) 308-311. https://doi.org/10.1021/cm0011559

[25] M. Vallet-Regí, M Colilla, I Izquierdo-Barba, J. Biomed. Nanotechnol. 4 (2008) 1-15. https://doi.org/10.1166/jbn.2008.002

[26] M. Vallet-Regí, M. Colilla, I. Izquierdo-Barba, M. Manzano, Molecules 23 (2018) 47. https://doi.org/10.3390/molecules23010047

[27] G. Villaverde, A. Alfranca, A. González-Murillo, G. J. Melen, R. R. Castillo, M. Ramírez, A. Baeza, M. Vallet-Regí, Angew. Chem. Int. Ed. 58 (2019) 3067-3072. https://doi.org/10.1002/anie.201811691

[28] P. Mora-Raimundo, D. Lozano, M. Manzano, M. Vallet-Regí, ACS Nano 13 (2019) 5451-5464. https://doi.org/10.1021/acsnano.9b00241

[29] B. González, M. Colilla, J. Díez, D. Pedraza, M. Guembe, I. Izquierdo-Barba, M. Vallet-Regí, Acta Biomater. 68 (2018) 261-271. https://doi.org/10.1016/j.actbio.2017.12.041

[30] M. Martínez-Carmona, I. Izquierdo-Barba, M. Colilla, M. Vallet-Regí, Acta Biomater. 96 (2019) 547-556. https://doi.org/10.1016/j.actbio.2019.07.001

[31] R. R. Castillo, D. Lozano, B. González, M. Manzano, I. Izquierdo-Barba, M. Vallet-Regí, Expert Opin Drug Delivery 16 (2019) 415-439. https://doi.org/10.1080/17425247.2019.1598375

[32] M. N. Sanz-Ortiz, K. Sentosun, S. Bals, L. M. Liz-Marzán, ACS Nano 9 (2015) 10489-10497. https://doi.org/10.1021/acsnano.5b04744

[33] J. L. Paris, G. Villaverde, S. Gómez-Graña, M. Vallet-Regí, Acta Biomater. 101 (2020) 459-468. https://doi.org/10.1016/j.actbio.2019.11.004

[34] P. Liu, Y. Wang, Y. Liu, F. Tan, J. Li, N. Li, Theranostics 10 (2020) 6774-6789. DOI:10.7150/thno.42661

[35] J.S. Stamler, D.J. Singel, J. Loscalzo, Science 258 (1992) 1898-1902. DOI:10.1126/science.1281928

[36] A.W. Carpenter, M.H. Schoenfisch, Chem. Soc. Rev. 41 (2012) 3742-3752. https://doi.org/10.1039/C2CS15273H

[37] D.A. Riccio, M.H. Schoenfisch, Chem Soc. Rev. 41 (2012) 3731-3741. https://doi.org/10.1039/C2CS15272J

[38] J.F. Quinn, M.R. Whittaker, T.P. Davis, J. Control. Release 205 (2015) 190-205. https://doi.org/10.1016/j.jconrel.2015.02.007

[39] R. Guo, Y. Tian, Y. Wang, W. Yang, Adv. Funct. Mater. 27 (2017) 1606398. https://doi.org/10.1002/adfm.201606398

[40] Z. Sadrearhami, T.-K. Nguyen, R. Namivandi-Zangeneh, K. Jung, E.H.H. Wong, C. Boyer, J. Mater. Chem. B 6 (2018) 2945-2959. https://doi.org/10.1039/C8TB00299A

[41] E.M. Hetrick, J.H. Shin, N.A. Stasko, C.B. Johnson, D.A. Wespe, E. Holmuhamedov, M.H. Schoenfisch, ACS Nano 2 (2008) 235-246. https://doi.org/10.1021/nn700191f

[42] P.G. Wang, M. Xian, X. Tang, X. Wu, Z. Wen, T. Cai, A.J. Janczuk, Chem. Rev. 102 (2002) 1091-1134. https://doi.org/10.1021/cr000040l

[43] D.A. Riccio, J.L. Nugent, M.H. Schoenfisch, Chem. Mater. 23 (2011) 1727-1735. https://doi.org/10.1021/cm102510q

[44] K. Dong, E. Ju, N. Gao, Z. Wang, J. Ren, X. Qu, Chem. Commun. 52 (2016) 5312-5315. https://doi.org/10.1039/C6CC00774K

[45] R.N. Allan, M.J. Kelso, A. Rineh, N.R. Yepuri, M. Feelisch, O. Soren, S. Brito-Mutunayagam, R.J. Salib, P. Stoodley, S.C. Clarke, J.S. Webb, L. Hall-Stoodley, S.N. Faust, Nitric Oxide 65 (2017) 43-49. https://doi.org/10.1016/j.niox.2017.02.006

[46] A. Gole, C.J. Murphy, Chem. Mater. 16 (2004) 3633-3640. https://doi.org/10.1021/cm0492336

[47] L. Scarabelli, M. Grzelczak, L.M. Liz-Marzán, Chem. Mater. 25 (2013) 4232-4238. https://doi.org/10.1021/cm402177b

[48] G. Villaverde, S. Gómez-Graña, E. Guisasola, I. García, C. Hanske, L.M. Liz-Marzán, A. Baeza, M. Vallet-Regí, Part. Part. Syst. Charact. 35 (2018) 1800148. https://doi.org/10.1002/ppsc.201800148

[49] B. Herigstad, M. Hamilton, J. Heersink, J. Microbiol. Methods 44 (2001) 121-129. https://doi.org/10.1016/S0167-7012(00)00241-4

[50] M.A. Hamilton, A.E. Parker. Enumerating viable cells by pooling counts for several dilutions. © 2010 MSU Center of Biofilm Engineering.

[51] N.S. Abadeer, M.R. Brennan, W.L. Wilson, C.J. Murphy, ACS Nano 8 (2014) 8392-8406. https://doi.org/10.1021/nn502887j

[52] F. Hoffmann, M. Cornelius, J. Morell, M. Fröba, Angew. Chem. Int. Ed. 45 (2006) 3216-3251. https://doi.org/10.1002/anie.200503075

[53] M.H. Lim, A. Stein, Chem. Mater. 11 (1999) 3285-3295. https://doi.org/10.1021/cm990369r

[54] F. de Juan, E. Ruiz-Hitzky, Adv. Mater. 12 (2000) 430-432. https://doi.org/10.1002/(SICI)1521-4095(200003)12:6<430::AID-ADMA430>3.0.CO;2-3

[55] J. Kecht, A. Schlossbauer, T. Bein, Chem. Mater. 20 (2008) 7207-7214. https://doi.org/10.1021/cm801484r

[56] Á. Martínez, E. Fuentes-Paniagua, A. Baeza, J. Sánchez-Nieves, M. Cicuéndez, R. Gómez, F.J. de la Mata, B. González, M. Vallet-Regí, Chem. Eur. J. 21 (2015) 15651-15666. https://doi.org/10.1002/chem.201501966

[57] L.T. Zhuravlev, Colloids Surf. A (2000), 173, 1-38. https://doi.org/10.1016/S0927-7757(00)00556-2

[58] W. Fan, W. Bu, Z. Zhang, B. Shen, H. Zhang, Q. He, D. Ni, Z. Cui, K. Zhao, J. Bu, J. Du, J. Liu, J. Shi, Angew. Chem. Int. Ed. 54 (2015) 14026-14030. https://doi.org/10.1002/anie.201504536

[59] A. Lutzke, J.B. Tapia, M.J. Neufeld, M.M. Reynolds. ACS Appl. Mater. Interfaces 9 (2017) 2104-2113. https://doi.org/10.1021/acsami.6b12888

[60] A. Nieto, M. Colilla, F. Balas, M. Vallet-Regí, Surface electrochemistry of mesoporous silicas as a key factor in the design of tailored delivery devices, Langmuir 26 (2010) 5038-5049. https://doi.org/10.1021/la904820k

[61] The Genetic Code and the Origin of Life. Ed. L. Ribas de Pouplana. Springer Nature, 2004. Ribozyme-Catalyzed Genetics. D. H. Burke, Chapter 4, pp. 48-74.

[62] Suriyanto, E.Y.K. Ng, S.D. Kumar. Biomed. Eng. OnLine 16 (2017) 36. https://doi.org/10.1186/s12938-017-0327-x

[63] M. Cicuéndez, I. Izquierdo-Barba, M.T. Portolés, M. Vallet-Regí, Biocompatibility and levofloxacin delivery of mesoporous materials, Eur. J. Pharm. Biopharm. 84 (2013) 115-124. DOI:10.1016/j.ejpb.2012.11.029

[64] M. Cicuéndez, J.C. Doadrio, A. Hernández, M.T. Portolés, I. Izquierdo-Barba, M. Vallet-Regí, Multifunctional pH sensitive 3D scaffolds for treatment and prevention of bone infection, Acta Biomaterialia 65 (2018) 450-461. https://doi.org/10.1016/j.actbio.2017.11.009

**Tables**

**Table 1.** Characterization of the different AuNR@MSN nanosystems by ζ-potential and hydrodynamic particle size (maximum of the distribution with population percentages) by DLS.

| Sample | ζ Potential (mV) | Size (nm) |
|---|---|---|
| **AuNR@MSN** | −21.2 ± 1.0 | 68.1 ± 14.1 (17.7%) |
| **AuNR@MSN-PEG$_{ext}$** | −16.9 ± 0.7 | 91.3 ± 7.9 (19.3%) |
| **AuNR@MSN-SH** | −20.7 ± 0.5 | 78.8 ± 12.7 (19.2%) |
| **AuNR@MSN-SNO** | −12.6 ± 0.2 | 105.7 ± 16.1 (18.2%) |

**Table 2.** Elemental chemical analysis (atomic percentages) of the different AuNR@MSN nanosystems.

| Sample | C (%) | H (%) | N (%) | S (%) |
|---|---|---|---|---|
| **AuNR@MSN** | 4.65 | 2.22 | 0.14 | 0.01 |
| **AuNR@MSN-PEG$_{ext}$** | 9.26 | 2.30 | 0.21 | 0.10 |
| **AuNR@MSN-PEG$_{ext}$+LEVO** | 9.32 | 2.09 | 0.52 | 0.08 |
| **AuNR@MSN-SH** | 7.41 | 1.99 | 0.15 | 1.81 |
| **AuNR@MSN-SNO** | 7.56 | 1.98 | 0.12 | 1.69 |

**Figures**

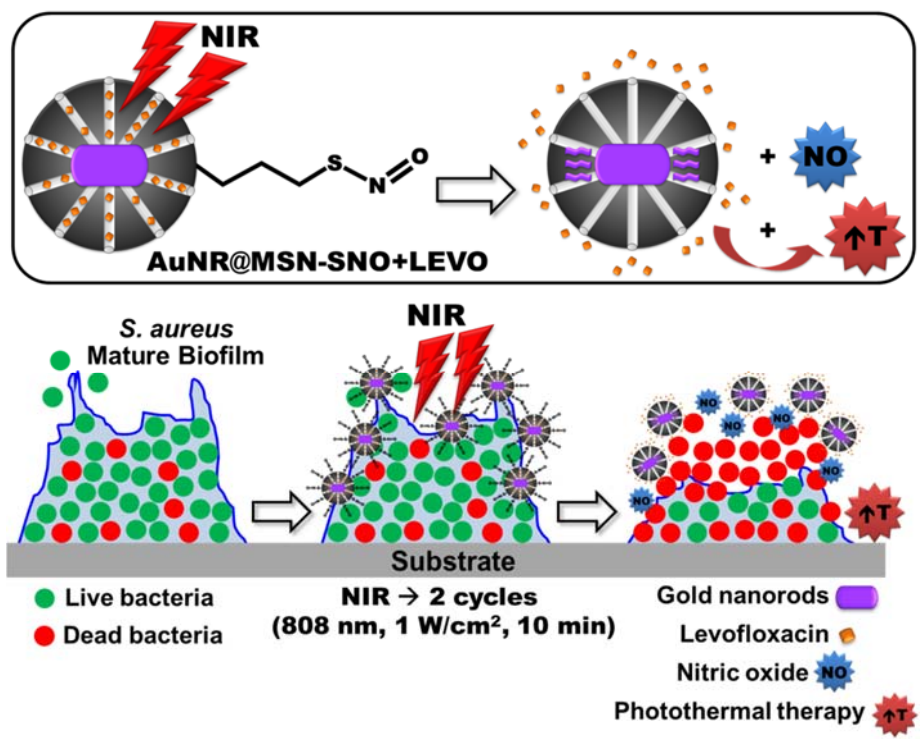

**Figure 1.** Schematic representation of the nanosystem design AuNR@MSN-SNO+LEVO and its effect on a *S. aureus* biofilm in response to NIR laser treatment.

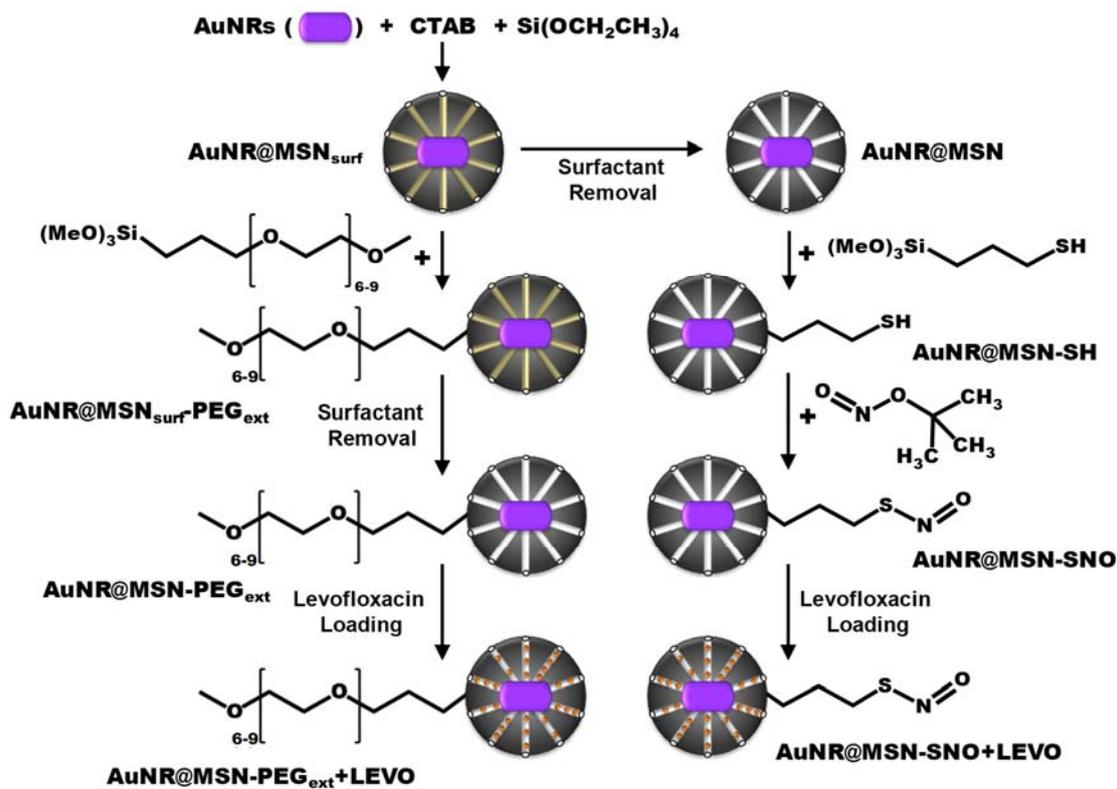

**Figure 2.** Synthesis of the different AuNR@MSN nanosystems.

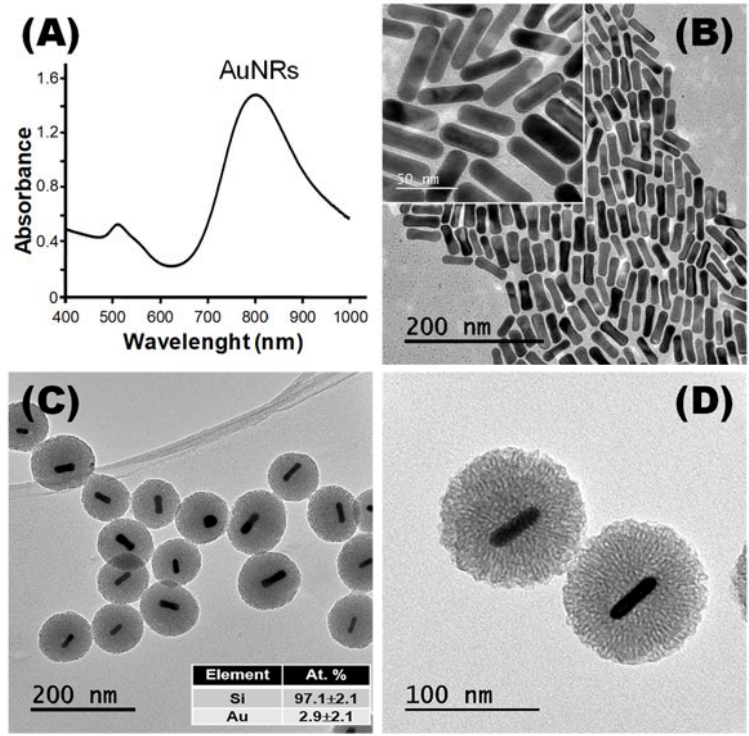

**Figure 3.** UV-Visible absorption spectrum (A) and TEM micrographs (B, the inset shows an image at higher magnification) of synthetized gold nanorods (AuNRs). TEM micrographs of AuNRs surrounded by a mesoporous silica shell AuNR@MSN$_{surf}$ (C) and AuNR@MSN-PEG$_{ext}$ (D) nanosystems. Atomic percentage obtained by EDS analysis for Si and Au in AuNR@MSN nanosystems was included as an inset (C).

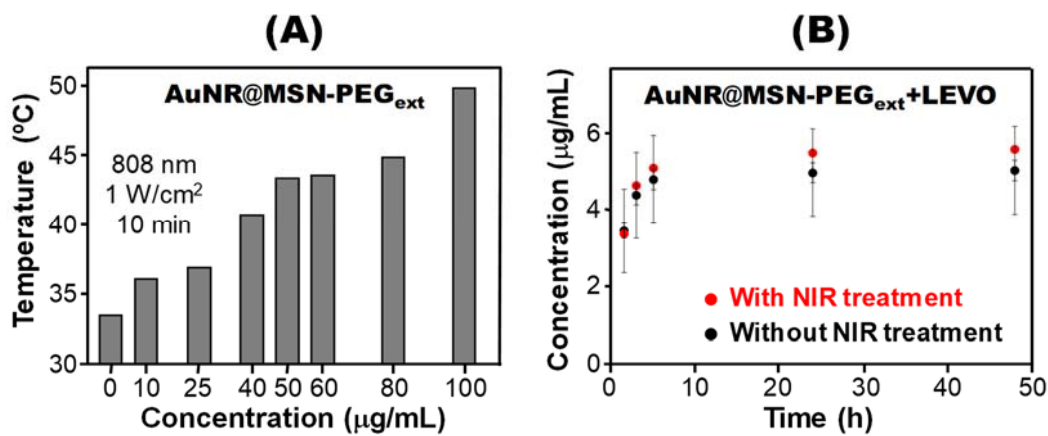

**Figure 4.** (A) Temperature increase of the culture media after NIR treatment as a function of AuNR@MSN-PEG$_{ext}$ concentration. (B) Levofloxacin release curve for AuNR@MSN-PEG$_{ext}$+LEVO with or without NIR treatment. Data are expressed as cumulative concentration. Experiments were performed on glass cover substrates using a wavelength of 808 nm at a power of 1 W/cm$^2$ for a NIR irradiation time of 10 minutes.

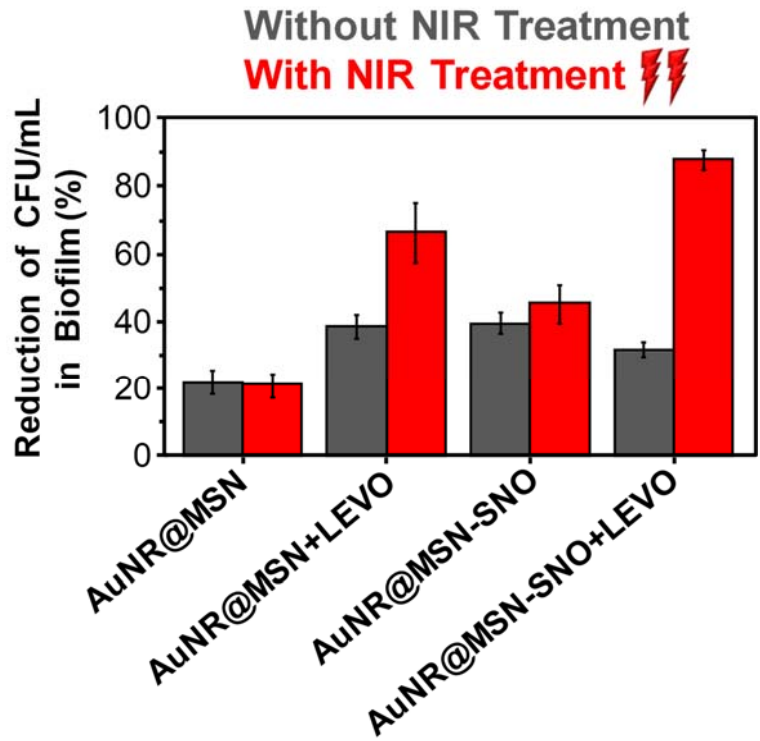

**Figure 5.** Percentage reduction of cell viability (CFU/mL) of *S. aureus* biofilm by the action of the different AuNR@MSN nanosystems without (grey) and with (red) two cycles of NIR laser treatment (808 nm, 1 W/cm$^2$, 10 min).